

\font\titlefont = cmr10 scaled\magstep 4
\font\sectionfont = cmr10
\font\littlefont = cmr5
\font\eightrm = cmr8

\def\sss{\scriptscriptstyle}

\magnification = 1200

\global\baselineskip = 1.2\baselineskip
\global\parskip = 4pt plus 0.3pt
\global\abovedisplayskip = 18pt plus3pt minus9pt
\global\belowdisplayskip = 18pt plus3pt minus9pt
\global\abovedisplayshortskip = 6pt plus3pt
\global\belowdisplayshortskip = 6pt plus3pt


\def\endignore{}
\def\ignore #1\endignore{}

\newcount\dflag
\dflag = 0


\def\monthname{\ifcase\month
\or Jan \or Feb \or Mar \or Apr \or May \or June%
\or July \or Aug \or Sept \or Oct \or Nov \or Dec
\fi}

\def\timestring{{\count0 = \time%
\divide\count0 by 60%
\count2 = \count0
\count4 = \time%
\multiply\count0 by 60%
\advance\count4 by -\count0
\ifnum\count4 < 10 \toks1 = {0}
\else \toks1 = {} \fi%
\ifnum\count2 < 12 \toks0 = {a.m.}
\else \toks0 = {p.m.}
\advance\count2 by -12%
\fi%
\ifnum\count2 = 0 \count2 = 12 \fi
\number\count2 : \the\toks1 \number\count4%
\thinspace \the\toks0}}

\def\today{\ifcase\month\or January\or February\or March\or
 April\or May\or June\or July\or August\or September\or
 October\or November\or December\fi \space\number\day, \number\year}



\def\endtitle{}
\def\title#1\endtitle{\vskip.5in\titlefont
\global\baselineskip = 2\baselineskip
#1\vskip.4in
\baselineskip = 0.5\baselineskip\rm}

\def\endauthors{}
\def\authors#1\endauthors{#1}

\def\endabstract{}
\def\abstract#1\endabstract{\vskip .3in%
\centerline{\sectionfont\bf Abstract}%
\vskip .1in
\noindent#1}

\newcount\nsection
\newcount\nsubsection

\def\section#1{\global\advance\nsection by 1
\nsubsection=0
\bigskip\noindent\centerline{\sectionfont \bf \number\nsection.\ #1}
\bigskip\rm\nobreak}

\def\subsection#1{\global\advance\nsubsection by 1
\bigskip\noindent\sectionfont \sl \number\nsection.\number\nsubsection)\
#1\bigskip\rm\nobreak}

\def\topic#1{{\medskip\noindent $\bullet$ \it #1:}}
\def\endtopic{\medskip}

\def\appendix#1#2{\bigskip\noindent%
\centerline{\sectionfont \bf Appendix #1.\ #2}
\bigskip\rm\nobreak}


\newcount\nref
\global\nref = 1

\def\ref#1#2{\xdef #1{[\number\nref]}
\ifnum\nref = 1\global\xdef\therefs{\noindent[\number\nref] #2\ }
\else
\global\xdef\oldrefs{\therefs}
\global\xdef\therefs{\oldrefs\vskip.1in\noindent[\number\nref] #2\ }%
\fi%
\global\advance\nref by 1
}

\def\listrefs{\vfill\eject\section{References}\therefs}


\newcount\nfoot
\global\nfoot = 1

\def\foot#1#2{\xdef #1{(\number\nfoot)}
\footnote{${}^{\number\nfoot}$}{\eightrm #2}
\global\advance\nfoot by 1
}


\newcount\nfig
\global\nfig = 1

\def\fig#1#2{\xdef #1{(\number\nfig)}
\global\advance\nfig by 1
}


\newcount\cflag
\newcount\nequation
\global\nequation = 1
\def\eqlabel{(1)}

\def\nexteqno{\ifnum\cflag = 0
\global\advance\nequation by 1
\fi
\global\cflag = 0
\xdef\eqlabel{(\number\nequation)}}

\def\lasteqno{\global\advance\nequation by -1
\xdef\eqlabel{(\number\nequation)}}

\def\label#1{\xdef #1{(\number\nequation)}
\ifnum\dflag = 1
{\escapechar = -1
\xdef\draftname{\littlefont\string#1}}
\fi}

\def\clabel#1#2{\xdef\eqlabel{(\number\nequation #2)}
\global\cflag = 1
\xdef #1{\eqlabel}
\ifnum\dflag = 1
{\escapechar = -1
\xdef\draftname{\string#1}}
\fi}

\def\cclabel#1#2{\xdef\eqlabel{#2)}
\global\cflag = 1
\xdef #1{\eqlabel}
\ifnum\dflag = 1
{\escapechar = -1
\xdef\draftname{\string#1}}
\fi}


\def\eeq{}

\def\eqnn #1\eeq{$$ #1 $$}

\def\eq #1\eeq{\xdef\draftname{\ }
$$ #1
\eqno{\eqlabel \rlap{\ \draftname}} $$
\nexteqno}



\def\eol{& \eqlabel \rlap{\ \draftname} \crcr
\nexteqno
\xdef\draftname{\ }}

\def\eeol{& \eqlabel \rlap{\ \draftname}
\nexteqno
\xdef\draftname{\ }}

\def\eolnn{\cr
\global\cflag = 0
\xdef\draftname{\ }}

\def\eeolnn{\xdef\draftname{\ }}

\def\eqa #1\eeq{\xdef\draftname{\ }
$$ \eqalignno{ #1 } $$
\global\cflag = 0}


\def\ie{{\it i.e.\/}}
\def\eg{{\it e.g.\/}}
\def\etc{{\it etc.\/}}

\def\apriori{{\it a priori\/}}

\def\via{{\it via\/}}
\def\vs{{\it vs.\/}}
\def\cf{{\it c.f.\/}}


\def\anp#1#2#3{{\it Ann.~Phys. (NY)} {\bf #1} (19#2) #3}
\def\arnps#1#2#3{{\it Ann.~Rev.~Nucl.~Part.~Sci.} {\bf #1}, (19#2) #3}

\def\ijmp#1#2#3{{\it Int.~J.~Mod.~Phys.} {\bf A#1} (19#2) #3}

\def\mpla#1#2#3{{\it Mod.~Phys.~Lett.} {\bf A#1}, (19#2) #3}

\def\npb#1#2#3{{\it Nucl.~Phys.} {\bf B#1} (19#2) #3}
\def\plb#1#2#3{{\it Phys.~Lett.} {\bf #1B} (19#2) #3}
\def\prc#1#2#3{{\it Phys.~Rev.} {\bf C#1} (19#2) #3}
\def\prd#1#2#3{{\it Phys.~Rev.} {\bf D#1} (19#2) #3}
\def\pr#1#2#3{{\it Phys.~Rev.} {\bf #1} (19#2) #3}

\def\prl#1#2#3{{\it Phys.~Rev.~Lett.} {\bf #1} (19#2) #3}

\def\zpc#1#2#3{{\it Zeit.~Phys.} {\bf C#1} (19#2) #3}


\global\nulldelimiterspace = 0pt



\def\frac#1#2{{{#1} \over {#2}}\,}  
\def\hf{{1\over 2}}



\def\Dsl{\hbox{/\kern-.6700em\it D}} 
\def\dsl{\hbox{/\kern-.5300em$\partial$}}
\def\pxpsl{\hbox{/\kern-.5600em$p$}}
\def\ssl{\hbox{/\kern-.5300em$s$}}
\def\epssl{\hbox{/\kern-.5100em$\epsilon$}}
\def\delsl{\hbox{/\kern-.6300em$\nabla$}}
\def\lxpsl{\hbox{/\kern-.4300em$l$}}
\def\elxpsl{\hbox{/\kern-.4500em$\ell$}}
\def\kxpsl{\hbox{/\kern-.5100em$k$}}
\def\qxpsl{\hbox{/\kern-.5000em$q$}}
\def\sla#1{\raise.15ex\hbox{$/$}\kern-.57em #1}



\def\roughly#1{\mathrel{\raise.3ex\hbox{$#1$\kern-.75em\lower1ex\hbox{$\sim$}}}}
\def\lsim{\roughly<}
\def\gsim{\roughly>}

\def\ol#1{\overline{#1}}




\def\Bfw{{\bf W}}


\def\Sca{{\cal A}}

\def\Scd{{\cal D}}

\def\Scl{{\cal L}}

\def\Sco{{\cal O}}

\def\Scw{{\cal W}}

\def\Scz{{\cal Z}}


\def\tr{\mathop{\rm tr}}
\def\Tr{\mathop{\rm Tr}}

\def\Det{\mathop{\rm Det}}
\def\Log{\mathop{\rm Log}}







\def\gwk{$SU_L(2) \times U_Y(1)$}
\def\gem{$U_{\rm em}(1)$}
\def\em{{\rm em}}
\def\ngb{Nambu-Goldstone boson}
\def\zdm{$Z$dm}

\def\sq{{\vbox {\hrule height 0.6pt\hbox{\vrule width 0.6pt\hskip 3pt
        \vbox{\vskip 6pt}\hskip 3pt \vrule width 0.6pt}\hrule height 0.6pt}}}
\def\mw{M_{\sss W}}
\def\mz{M_{\sss Z}}
\def\mh{m_{\sss H}}

\def\leff{{\Scl_{\rm eff}}}


\rightline{March 1992; revised June 1993}
\rightline{McGill-92/05}
\rightline{UdeM-LPN-TH-84}
\rightline{hepph/9203216}
\vskip .3in

\title
\centerline{Uses and Abuses of}
\centerline{Effective Lagrangians}
\endtitle

\authors
\centerline{C.P.~Burgess${}^a$ and David London${}^b$
\footnote{}{email:
cliff@physics.mcgill.ca; london@lps.umontreal.ca}}
\vskip .15in
\centerline{\it ${}^a$ Physics Department, McGill University}
\centerline{\it 3600 University St., Montr\'eal, Qu\'ebec, CANADA, H3A 2T8.}
\vskip .1in
\centerline{\it ${}^b$ Laboratoire de Physique Nucl\'eaire, Universit\'e de
Montr\'eal}
\centerline{\it C.P.~6128, Montr\'eal, Qu\'ebec, CANADA, H3C 3J7.}
\endauthors

\abstract
Motivated by past and recent analyses we critically re-examine the use of
effective lagrangians in the literature to constrain new physics and to
determine the `physics reach' of future experiments. We demonstrate that
many calculations, such as those involving anomalous trilinear gauge-boson
couplings, either considerably overestimate loop-induced effects, or give
ambiguous answers. The source of these problems is the use of cutoffs to
evaluate the size of such operators in loop diagrams. In contrast to other
critics of these loop estimates, we prove that the inclusion of
nonlinearly-realized gauge invariance into the low-energy lagrangian is
irrelevant to this conclusion. We use an explicit example using known
multi-Higgs physics above the weak scale to underline these points. We show
how to draw conclusions regarding the nature of the unknown high-energy
physics without making reference to low-energy cutoffs.
\endabstract

\vfill\eject


\section{Introduction}

As experimentally accessible energies have risen above the thresholds for
producing electroweak gauge bosons it has become more and more clear that
the mass scale associated with any new physics is probably at significantly
higher energies. This is reflected by the great success of the standard
model in predicting the results of these experiments in general and the
properties of these gauge bosons in particular.

Given that the scale of physics beyond the standard model is well above the
weak scale, the low-energy effects of such new physics may be parametrized
in terms of an effective lagrangian
\ref\effective{S. Weinberg, Physica {\bf 96A} (1979) 327;
J. Polchinski, \npb{231}{84}{269};
H. Georgi, {\it Weak Interactions and Modern Particle Theory}
(Benjamin/Cummings Menlo Park, 1984);
C.P. Burgess and J.A. Robinson, in {\it BNL Summer Study on CP Violation}
S. Dawson and A. Soni editors, (World Scientific, Singapore, 1991).}
\effective\ in which the influence of any at-present-unknown new heavy
particles is felt through the effective nonrenormalizable interactions that
they generate among the lighter particles. These nonstandard interactions
may be organized according to increasing operator dimension. At a practical
level this method is useful only to the extent that it is possible to
consider just those few interactions which have the lowest dimension. This
can usually be justified by the suppression of higher-dimension operators
by extra powers of the inverse of some heavy mass scale, $M$.

This type of reasoning has led to considerable effort in using experimental
data to constrain the coefficients of the operators in such an effective
lagrangian which parametrize deviations from the standard model. Of
particular interest are those terms which correspond to anomalous couplings
of the photon and the $Z^0$, since these are the probes that are currently
the most cleanly available in collider experiments. Analyses have focused
on the lowest electromagnetic and electroweak moments of the light fermions
\ref\fermionsnoninv{W. Bernreuther and O. Nachtmann, \prl{63}{89}{2787};
W. Bernreuther, U. L\"ow, J.P. Ma and O. Nachtmann, \zpc{43}{89}{117};
G. Valencia and A. Soni, \plb{263}{91}{517}.}
\ref\fermionsnonlin{R.D. Peccei, S. Peris and X. Zhang,
\npb{349}{91}{305}.}
\ref\fermionslin{W. Buchm\"uller and D. Wyler, \npb{268}{86}{621};
J. Anglin, C.P. Burgess, H. de Guise, C. Mangin and J.A. Robinson,
\prd{43}{91}{703};
C.P. Burgess and J.A. Robinson, \ijmp{6}{91}{2707};
A. De R\'ujula, M.B. Gavela, O. P\`ene and F.J. Vegas, \npb{357}{91}{311};
M. Traseira and F.J. Vegas, \plb{262}{91}{12}.}
\fermionsnoninv, \fermionsnonlin, \fermionslin\ as well as gauge-boson
self-couplings
\ref\wwv{K.J.F. Gaemers and G.J. Gounaris, \zpc{1}{79}{259};
K. Hagiwara, R.D. Peccei, D. Zeppenfeld and K. Hikasa, \npb{282}{87}{253}.}
\ref\gbtreenoninv{ C.L. Bilchak and J.D. Stroughair, \prd{30}{84}{1881};
J.A. Robinson and T.G. Rizzo, \prd{33}{86}{2608};
S.-C. Lee and W.-C. Su, \prd{38}{88}{2305}; \plb{205}{88}{569};
\plb{212}{88}{113}; \plb{214}{88}{276};
C.-H. Chang and S.-C. Lee, \prd{37}{88}{101};
D. Zeppenfeld and S. Willenbrock, \prd{37}{88}{1775};
U. Baur and D. Zeppenfeld, \npb{308}{88}{127}; \plb{201}{88}{383};
\npb{325}{89}{253}; \prd{41}{90}{1476};
G. Couture, S. Godfrey and P. Kalyniak, \prd{39}{89}{3239};
K. Hagiwara, J. Woodside and D. Zeppenfeld, \prd{41}{90}{2113};
V. Barger and T. Han, \plb{241}{90}{127};
E.N. Argyres, O. Korakianitis, C.G. Papadopoulos and W.J. Stirling,
\plb{259}{91}{195};
E.N. Argyres and C.G. Papadopoulos, \plb{263}{91}{298}; S. Godfrey, Carleton
preprint OCIP-C-91-2;
G. Couture, S. Godfrey and R. Lewis, \prd{45}{92}{777}.}
\ref\wdominance{
P.Q. Hung and J.J. Sakurai, \npb{143}{78}{81}; J.D. Bjorken, \prd{19}{79}{335};
R. K\"ogerler and D. Schildknecht, CERN preprint CERN-TH.3231 (1982),
unpublished;  D. Schildknecht, in ``Electroweak effects at high energies'', ed.
N.B. Newman (Plenum, New York, 1985) p.~551;
M. Kuroda, D. Schildknecht and K.-H. Schwarzer, \npb{261}{85}{432};
J. Maalampi, D. Schildknecht, and K.-H. Schwarzer, \plb{166}{86}{361};
M. Kuroda, J. Maalampi, D. Schildknecht and K.-H. Schwarzer,
\plb{190}{87}{217}; \npb{284}{87}{271};
H. Neufeld, J.D. Stroughair and D. Schildknecht, \plb{198}{87}{563};
M. Kuroda, F.M. Renard and D. Schildknecht, \plb{183}{87}{366};
C. Bilchak, M. Kuroda and D. Schildknecht, \npb{299}{88}{7};
Y. Nir, \plb{209}{88}{523};
H. Schlereth, \plb{256}{91}{267}.}
\ref\gbloopnoninv{S.J. Brodsky and J.D. Sullivan, \pr{156}{67}{1644};
F. Herzog, \plb{148}{84}{355}; \plb{155}{85}{468}E;
J.C. Wallet, \prd{32}{85}{813};
A. Grau and J.A. Grifols, \plb{154}{85}{283};
F. Hoogeveen, Max Planck Inst.~preprint MPI.PAE/PTh 25/87 (1987), unpublished.
W.J. Marciano and A. Queijeiro, \prd{33}{86}{3449};
G. B\'elanger, F. Boudjema and D. London, \prl{665}{90}{2943};
F. Boudjema, K. Hagiwara, C. Hamzaoui and K. Numata, \prd{43}{91}{2223}.}
\ref\gblooplin{D. Atwood, C.P. Burgess, C. Hamzaoui, B. Irwin and J.A.
Robinson,
\prd{42}{90}{3770}; F. Boudjema, C.P. Burgess, C. Hamzaoui and J.A. Robinson,
\prd{43}{91}{3683}.}
\ref\errors{
M. Suzuki, \plb{153}{85}{289};
A. Grau and J.A. Grifols, \plb{166}{86}{233};
J.J. van der Bij, \prd{35}{87}{1088};
G.L. Kane, J. Vidal and C.-P. Yuan, \prd{39}{89}{2617};
H. Neufeld, J.D. Stroughair and D. Schildknecht, Ref.~\wdominance;
Y. Nir, Ref.~\wdominance;
J.A. Grifols, S. Peris and J. Sol\`a, \plb{197}{87}{437}; \ijmp{3}{88}{225};
R. Alcorta, J.A. Grifols and S. Peris, \mpla{2}{87}{23};
C. Bilchak and J.D. Stroughair, \prd{41}{90}{2233};
P. M\'ery, S.E. Moubarik, M. Perrottet and F.M. Renard, \zpc{46}{90}{229};
R.D. Peccei and S. Peris, \prd{44}{91}{809};
H. K\"onig, \prd{45}{92}{1575};
D. London, \prd{45}{92}{3186};
S. Godfrey and H. K\"onig, \prd{45}{92}{3196}.}
\ref\criticisms{A. de R\'ujula, M.B. Gavela, P. Hernandez and E. Mass\'o,
\npb{384}{92}{3}.}
\wwv, \gbtreenoninv, \wdominance, \gbloopnoninv, \gblooplin, \errors,
\criticisms\ that would dominate interactions at low energies. In this way
it is possible to ascertain which interactions could have hitherto escaped
detection and might yet be detectable at upcoming experiments. Proponents
of particular experiments can turn this argument around and estimate the
scale, $M$, of new physics to which a particular proposal can be
sensitive---its so-called `physics reach'. The most interesting proposals
are naturally those that are potentially sensitive to the highest scales
and so whose physics reach is the longest.

So far so good. A complication arises, however, when loop effects in the
low-energy theory are important for detecting the effective interaction
under study. This is because such loops are typically divergent and so can
depend on {\it positive} powers of a large high-energy cutoff, $\Lambda$.
This cutoff physically describes the maximum energy to which the effective
lagrangian is expected to apply and so is frequently also taken to be of
order of the new physics scale, $M$. To the extent that this is true the
most divergent contributions to a given amplitude could be taken as
indications of a strong dependence on new physics at scale $M$, potentially
indicating a long physics reach.

Our main point in this paper is to show that the above argument can be very
misleading, and can even lead to conclusions which contradict general
decoupling results
\ref\actheorem{T. Appelquist and J. Carrazone, \prd{11}{75}{2856};
J. Polchinski, \npb{231}{84}{269};
B.J. Warr, \anp{183}{88}{1}; \anp{183}{88}{59}.}
\actheorem. At best it gives \errors\ an ambiguous --- and at worst, a
false --- indication of the scale of new physics to which a given
experiment may be sensitive, often yielding overly stringent constraints on
parameters in the effective lagrangian. The weak link in the arguments used
is the assumed connection between what can be computed (the
cutoff dependence of amplitudes in the low-energy effective theory) and
what is meant to be bounded (the dependence of low-energy amplitudes on
physical high-energy physics scales such as heavy particle masses).

In this paper we refine and expand on our results in
\ref\physrevlet{C.P. Burgess and David London, \prl{69}{92}{3428}.}
Ref.~\physrevlet, by exploring in detail this connection between low-energy
cutoff dependence and heavy-mass dependence. We demonstrate our conclusions
within the context of a multi-Higgs model in which the influence of the
high-energy physics is known and calculable. We show that cutoff dependence
can be a very poor indicator of heavy-mass dependence, particularly where
massive spin-one particles are involved. We then indicate how to extract
the dependence on high-frequency physics without resorting to arguments
that rely on cutoffs.

In the literature, the misidentification of heavy-mass and cutoff
dependence arises most frequently in the context of anomalous
three-gauge-boson vertices (TGV's). There are two reasons for this. First,
since TGV's cannot yet be measured directly, the only available information
concerning them arises indirectly through their contributions to loops.
Second, since problems with interpreting cutoff dependence arise most
strikingly for loops involving massive spin-one particles, TGV-induced
loops are very easy to mishandle \errors. This has led to misleadingly
stringent constraints on anomalous TGV's, as well as to mistaken
predictions of large effects in future experiments -- that is to say: long
physics reach.

In addition to this confusion between cutoff behaviour and new-physics
dependence, the waters have recently become even more muddied due to a
parallel confusion that has arisen within the specific context of TGV
analyses. The authors of Ref.~\criticisms\ agree that physics reach as
regards anomalous TGV's is overstated in places in the literature. However,
they go on to identify the error as being the gauge invariance (or lack
thereof) of the analysis. (An alternative phrasing of this line of thought
is to object to the use of unitary gauge in performing loop calculations.)

The key question is whether the light particles in the effective theory
being considered fill out a {\it linear} representation of the gauge group.
They do not, for instance, if there is no light Higgs boson to transform
with the longitudinal $W$ and $Z$ bosons. In this case, gauge invariance
can only be realized nonlinearly. We contend here that, for gauge
symmetries, such a nonlinear realization can be included, or not, simply by
a change of variables, and so nothing physical can depend on this choice.

That this confusion can arise at all serves to underline a more pervasive
hazard that underlies the association of a physical interpretation to
divergences within an effective lagrangian: the lagrangians themselves, and
so also the divergences they contain, are {\it not} invariant under field
redefinitions. Conclusions that are based on them are generically marked by
the same flaw, unless it is specifically demonstrated otherwise (as can be
done for the $S$-matrix, for example). Proposals which link
cutoff dependence in the lagrangian to heavy-mass dependence are therefore
at best ambiguous, unless they are specifically referred to a set of
variables which are to be used. They are simply wrong if the variables used
are poorly chosen.

Some of these points are undoubtedly familiar to some of the
effective-lagrangian cognicenti. They have not, however, been absorbed into
the wider community which is now finding applications for these techniques.
We therefore feel that an examination of the issues is timely given the
present debate over the accuracy of estimates of physics reach, and over
the nature of the properties that should be built into low-energy
lagrangians.

We next expose all of these points in more detail, with reference to
explicit underlying models for which both heavy-mass dependence and cutoff
dependence are separately calculable. We start in section (2) by discussing
the relevance of gauge symmetries for effective lagrangians. In so doing we
(re)demonstrate the equivalence between nonlinearly-realized gauge
symmetries and no gauge symmetries at all. This is followed in section (3)
by some general observations about how cutoff dependence arises in
low-energy effective theories. Section (4) contains the guts of our
criticism. We first present the arguments for thinking that cutoffs might
track heavy masses, and then criticize these arguments. We provide several
examples which indicate how field redefinitions can alter cutoff
dependence, and argue which variables are most likely to allow cutoffs to
mimic heavy-mass dependence in observables. In section (5) we outline how
to infer heavy-mass dependence without having to rely on the cutoff
dependence of low-energy graphs. This permits the retention of most
applications of cutoff methods, but with the conceptual advantage of
relying on a more solid foundation. Section (6) then presents an explicit
multi-Higgs model for underlying physics in which these ideas are
explicitly worked out. Our conclusions are summarized in section (7).

\section{The Pertinence of Gauge Symmetries}

Essentially two ingredients are required to specify a low-energy effective
lagrangian: the low-energy particle content and the symmetries that their
interactions preserve. Once these have been specified, all possible
interactions of successively higher dimensions may generically be written
down.

When considering the interactions of $W$ and $Z$ bosons, the most important
distinction to be made concerns where the scale of the unknown new physics,
$M$, lies in relation to the electroweak scale, $v \simeq 246$ GeV. If $M$
is much greater than roughly $4 \pi v$, then the perturbative unitarity of
the low-energy theory requires that it must linearly realize the
electroweak gauge symmetries
\ref\cornwall{J.M. Cornwall, D.N. Levin and G. Tiktopoulos,
\prd{10}{74}{1145}.}
\ref\unitarity{B.W. Lee, C. Quigg and H. Thacker, \prd{16}{77}{1519};
M. Veltman, Acta.~Phys.~Pol. {\bf B8} (1977) 475.}
\cornwall, \unitarity. In this case the low-energy theory must contain more
particles than have presently been discovered (such as the standard-model
Higgs boson and top quark) in order for the known particles to fill out a
linear
representation of the gauge group. This is the choice that has been pursued
in Refs.~\fermionslin, \gblooplin, and \criticisms.

We are mostly concerned in what follows with the other alternative in which
the underlying physics we are groping for is the electroweak-breaking
physics itself. In this case the particle content need not fall into linear
representations of the gauge group, and so could in particular consist only
of those particles that have already been discovered. Since perturbative
unitarity fails in this type of effective theory at energies of order $4
\pi v \simeq 8 \pi \mw/g$, we are guaranteed that the effective theory must
fail at or before this point. Below this scale, agreement exists in the
literature as to the appropriate low-energy particle content that is to be
chosen, but practictioners divide according to their choices for the
symmetries that these particles should respect:

\topic{No Gauge Invariance}

In the first approach \fermionsnoninv, \wwv, \gbtreenoninv, \wdominance,
\gbloopnoninv, \errors, only electromagnetic gauge invariance is imposed,
and all spontaneously broken gauge symmetries are simply ignored.

\topic{Nonlinearly-Realized Gauge Invariance}
In the alternative framework \fermionsnonlin,
\ref\nonlinearrealization{T. Appelquist and C. Bernard, \prd{22}{80}{200};
A.C. Longhitano, Phys. Rev.~{\bf D22} (1980) 1166; \npb{188}{81}{118};
M.S. Chanowitz and M.K. Gaillard, \npb{261}{85}{379};
M.S. Chanowitz, M. Golden and H. Georgi, \prd{36}{87}{1490};
M.S. Chanowitz, \arnps{38}{88}{323};
R.D. Peccei and X. Zhang, \npb{337}{90}{269};
B. Holdom and J. Terning, \plb{247}{90}{88};
J. Bagger, S. Dawson and G. Valencia, preprint BNL-45782;
M. Golden and L. Randall, \npb{361}{91}{3};
B. Holdom, \plb{259}{91}{329};
A. Dobado, M.J. Herrero and D. Espriu, \plb{255}{91}{405};
R.D. Peccei and S. Peris, Ref.~\errors;
A. Dobado and M.J. Herrero, preprint CERN-TH-6272/91.}
\nonlinearrealization\ invariance with respect to the full electroweak
gauge group is required, but with all but the unbroken \gem\ subgroup being
nonlinearly realized. The physical motivation that underlies this second
approach is the assumption that the low-energy degrees of freedom of the
unknown symmetry-breaking sector contain only the three Nambu-Goldstone
bosons which are eaten by the massive $W$ and $Z$ particles. Given this
assumption, the transformation properties of all fields are determined by
general arguments
\ref\ccwz{S. Coleman, J. Wess and B. Zumino, \pr{177}{69}{2239};
E.C. Callan, S. Coleman, J. Wess and B. Zumino, \pr{177}{69}{2247};
J. Gasser and H. Leutwyler, \anp{158}{84}{142}.}
\ref\cgg{M.S. Chanowitz, M. Golden and H. Georgi, \prd{36}{87}{1490}.}
\ccwz, \cgg\ that were developed within the framework of chiral perturbation
theory many years ago.
\endtopic

It is the point of this section to (re)demonstrate the equivalence of these
last two schemes. This result is not new, appearing as it does in
Refs.~\cornwall\ and \cgg, but the reminder is worthwhile in order to put
to rest more recent concerns as to the legitimacy of ignoring the broken
electroweak symmetries in the effective lagrangian. The equivalence is
established by explicitly finding a change of variables that relates the
two alternatives. Although our arguments can be made quite generally, we
restrict ourselves here to two specific cases: a simplified toy model
involving a single massive spin-one particle, as well as the realistic case
appropriate to the couplings of the electroweak gauge bosons, $W^\pm$,
$Z^0$ and the photon, $\gamma$.

\subsection{The Toy Example}

In order to describe the argument within its simplest context, consider
first the coupling of a single massive spin-one particle, $V_\mu$, coupled
to various forms of spinless or spin-half matter, $\psi$. We first state
the two alternative forms for the effective lagrangian and then demonstrate
their equivalence.

\topic{No Gauge Invariance}
The lagrangian in the first formulation then takes the form:
\eq
\label\lagrangone
\Scl_1 = \Scl_1(V_\mu,\psi),
\eeq
in which $\Scl_1$ is \apriori\ an arbitrary local Lorentz-invariant
function of the fields $V_\mu$, $\psi$ and their spacetime derivatives.
Since $\psi$ and $V_\mu$ are independent degrees of freedom the quantum
theory could be defined in this case by a functional integral of the form:
\eq
\label\fintone
Z_1 = \int [d\psi] \; [dV_\mu] \; \exp\left[ i\int d^4x \; \Scl_1
(V_\mu,\psi) \right].
\eeq

\topic{Nonlinearly Realized Gauge Invariance}
The alternative formulation is to consider a $U(1)$ gauge theory with
matter fields, $\chi_i$, carrying $U(1)$ charges $q_i$. The gauge symmetry
transformations acting on these fields and on the gauge potential, $A_\mu$,
are the usual ones:
\eq
\chi_i \to e^{iq_i \omega} \; \chi_i; \qquad gA_\mu \to gA_\mu +
\partial_\mu \omega.
\eeq
$g$ here is the gauge coupling constant.

Symmetry breaking is incorporated by coupling these matter and gauge fields
in a completely general way to a single \ngb, $\varphi$, for a
spontaneously broken $U(1)$. The action of the $U(1)$ on the \ngb s may
always be chosen to take a standard form \ccwz, which becomes in this case
\eq
\varphi \to \varphi + f \omega.
\eeq
$f$ here is the \ngb's decay constant which is of the order of the scale at
which the $U(1)$ symmetry is spontaneously broken. It is related to the
mass  of the gauge boson by the relation $M = gf$.

The most general gauge-invariant low-energy lagrangian may then be written
in the following form:
\eq
\label\lagrangtwo
\Scl_2 = \Scl_2(D_\mu\varphi,\chi'),
\eeq
in which the redefined field is $\chi'_i \equiv e^{-i q_i \varphi/f} \;
\chi_i$ and the gauge-covariant derivative for $\varphi$ is given by $D_\mu
\varphi \equiv \partial_\mu \varphi - g f A_\mu$. Notice that all of the
dependence on $A_\mu$ in $\Scl_2$ arises through this gauge-covariant
derivative. For example, the gauge field strength is given by $gf F_{\mu\nu}
= \partial_\mu D_\nu \varphi - \partial_\nu D_\mu \varphi$.

The corresponding functional integral defining the quantum theory then has
the standard form:
\eq
\label\finttwo
Z_2 = \int [d\chi'_i] \; [dA_\mu] \; [d\varphi] \; \exp\left[ i\int d^4x \;
\Scl_2 (D_\mu\varphi,\chi') \right] \; \delta[G] \;
\Det\left(\frac{\delta G}{\delta \omega} \right),
\eeq
in which the second-to-last term is the functional delta function,
$\delta[G]$, which enforces the gauge condition $G = 0$, and the last term
is the associated Fadeev-Popov-DeWitt---or ghost---functional determinant.

It is crucial for the remainder of the argument that both $\chi'_i$ and
$D_\mu \varphi$ are {\it in}variant---as opposed to being {\it
co}variant---with respect to gauge transformations. As a result, {\it any}
Lorentz-invariant lagrangian, such as $\Scl_2$, that is built from these fields
becomes gauge invariant automatically.

\topic{Equivalence}
Now comes the main point. The two lagrangians, $\Scl_1$ and $\Scl_2$, are
identical to one another. There is a one-to-one correspondence between the
terms in each given by the replacement $\psi \leftrightarrow \chi'_i$ and
$D_\mu \varphi \leftrightarrow - gf \, V_\mu$. This is only possible
because {\it both} $\Scl_1$ and $\Scl_2$ are constrained only by Lorentz
invariance and so any interaction which is allowed for one is equally
allowed for the other.

More formally, the functional integral of eq.~\fintone\ may be obtained
from that of eq.~\finttwo\ by simply choosing unitary gauge, defined by the
condition $G \equiv \varphi(x)$, and using the functional delta function to
perform the integration over $\varphi$. The ghost `operator' is in this
case $\delta G(x)/\delta \omega(x') = f \; \delta^4(x-x')$ and so the ghost
determinant contributes just a trivial field-independent normalization
factor.

The integration over the `extra' Nambu-Goldstone degree of freedom of the
gauge-invariant theory is thereby seen to be precisely compensated by the
freedom to choose a gauge.

\subsection{Applications to the Electroweak Bosons}

The argument as applied to a more complicated symmetry-breaking pattern,
such as appears in the electroweak interactions, has essentially the same
logic although the technical details are slightly more intricate.

\topic{No Gauge Invariance}
We take for the purposes of illustration the degrees of freedom in the
low-energy effective lagrangian for the electroweak interactions of leptons
and quarks. These are: the massless photon, $A_\mu$, the massive weak
vector bosons, $W_\mu$ and $Z_\mu$, and the usual fermions, $\psi$.
Although other particles such as gluons may also be very simply included we
do not do so here for simplicity of notation. The general lagrangian for
these fields may be written:
\eq
\label\smlagrangone
\Scl_1 = \Scl_1(A_\mu,W_\mu, Z_\mu,\psi),
\eeq
in which $\Scl_1$ is a general local and Lorentz-invariant function whose
form is further constrained only by unbroken \gem-invariance. All
derivatives are taken to be the \gem\ gauge-covariant derivative, $D_\mu$,
which for fermions takes the form $D_\mu \psi = \partial_\mu \psi -ie Q
A_\mu \psi$. $Q$ here denotes the diagonal matrix of fermion electric
charges.

The quantum theory is given in terms of a functional integral of the form
\eq
\label\lowbrowint
Z_1 = \int [dW_\mu] \; [dW^*_\mu] \; [dZ_\mu] \; [dA_\mu] \;
[d\psi] \; \exp\left[ i\int d^4x \; \Scl_1 \right] \;
\delta\left[G_\em \right] \; \Det\left(\frac{\delta G_\em}{\delta
\omega_\em} \right).
\eeq

We next outline the nonlinear realization of \gwk.

\topic{Nonlinearly Realized Gauge Invariance}
The first step is to briefly review the formulation for the low-energy
interactions of the Nambu-Goldstone bosons for the {\it global}
symmetry-breaking pattern $SU_L(2) \times U_Y(1) \to U_\em(1)$ \ccwz. We
then promote the symmetry to local gauge transformations.

Consider, therefore, a collection of matter fields, $\psi$, on which \gwk\
is represented (usually reducibly) by the matrices $G = \exp[ i\omega_2^a
T_a + i \omega_1 Y]$. We choose here a slightly unconventional
normalization for the generators $T_a$ and $Y$, {\it viz} $\tr[ T_a T_b] =
\hf \, \delta_{ab}$, $[T_a Y] = 0$ and $\tr[Y^2] = \hf$. Finally define the
matrix-valued scalar field containing the Nambu-Goldstone bosons by $\xi(x)
= \exp[2i X_a \varphi^a(x) /v]$, in which the three $X_a$'s represent the
spontaneously broken generators $X_1 = T_1$, $X_2 = T_2$ and $X_3 = aT_3 -
bY$. Here $a^2+b^2=1$, and $a/b$ is chosen to ensure that $\tr[X_3 Q]=0$,
where $Q$ is the unbroken generator: $Q = b T_3 + a Y$.

The action of the gauge group \gwk\ on $\xi$ and $\psi$ may be written in
the standard form:
\eq
\label\realization
\psi \to G \psi \quad \hbox{and} \quad
\xi \to \xi', \quad \hbox{where} \quad G\; \xi = \xi' \; H^\dagger.
\eeq
Here $H = \exp[i Q \, u(\xi,\xi',G)]$ and $u = u(\xi,\xi',G)$ is implicitly
defined by the condition that $\xi'$ on the right-hand-side of
eq.~\realization\ involves only the broken generators.

As was the case for the toy example, for the purposes of constructing the
lagrangian it is convenient to define new matter fields, $\psi'$, according
to $\psi' \equiv \xi^\dagger \; \psi$ since this has the \gwk\
transformation rule:
\eqa
\psi' &\to \xi'^\dagger \; G \, \psi \eolnn
&= H \; \psi'.  \eeol
\eeq
Notice that even for global $U_Y(1)$ rotations, for which $\omega_1$ is
constant, $u(\xi,\xi',G)$ is spacetime dependent because of its dependence
on the scalar field $\xi(x)$.

The next step is the construction of the general locally \gwk\ invariant
effective lagrangian. To this end consider the auxiliary quantity
$\Scd_\mu(\xi)$ which may be defined in terms of $\xi$ and the \gwk\ gauge
potentials $\Bfw_\mu = g W^a_\mu \, T_a + g' B_\mu \, Y$ by
\eq
\label\auxiliary
\Scd_\mu(\xi) \equiv \xi^\dagger \partial_\mu\xi -i \xi^\dagger \Bfw_\mu \xi.
\eeq
In terms of this quantity it is possible to construct fields which
transform in a simple way with respect to \gwk. Together with their \gwk\
transformation rules these are,
\eqa
\label\connection
e\, \Sca_\mu & \equiv 2i \, \tr[ Q \Scd_\mu(\xi)], \qquad e\Sca_\mu \to
e\Sca_\mu + \partial_\mu u; \eolnn
\sqrt{{g'}^2 + g^2} \; \Scz_\mu &\equiv 2i \, \tr[X_3
\Scd_\mu(\xi)],\qquad \Scz_\mu \to \Scz_\mu; \eol
g \, \Scw^\pm_\mu &\equiv i\sqrt{2} \, \tr[T_\mp \Scd_\mu(\xi)], \qquad
\Scw^\pm_\mu \to e^{\pm iu Q} \, \Scw_\mu^\pm. \eeolnn
\eeq
$T_\pm$ is defined as usual to be $T_1 \pm iT_2$. The first of these
fields, $\Sca_\mu(\xi)$, transforms in such a way as to permit the
construction of a covariant derivative for the local transformations as
realized on $\psi'$:
\eq
\label\covderiv
D_\mu \psi' \equiv (\partial_\mu - i e\Sca_\mu \, Q )\; \psi'.
\eeq

The main point to be appreciated here is that eqs.~\connection\ imply that
all of the fields $\psi'$, $D_\mu\psi'$, $\Sca_\mu(\xi)$, $\Scz_\mu(\xi)$
and $\Scw_\mu^\pm(\xi)$ transform purely electromagnetically under
arbitrary \gwk\ transformations. This ensures that once the lagrangian is
constructed to be invariant under the unbroken group, \gem, it is {\it
automatically} invariant with respect to the full nonlinearly-realized
group \gwk.

With these transformation rules the most general \gwk-invariant lagrangian
becomes
\eq
\Scl_2 = \Scl_2(\Sca_\mu,\Scw_\mu,\Scz_\mu,\psi')
\eeq
with $\Scl_2$ restricted only by the unbroken \gem\ gauge invariance. The
functional integral which defines the quantum theory may then be written
\eq
\label\holygrailint
Z_2 = \int [d\Bfw_\mu] \; [d\xi] \; [d\psi'] \; \exp\left[
i\int d^4x \; \Scl_2 \right] \; \delta\left[G_a \right] \;
\Det\left(\frac{\delta G_a}{\delta \omega^b} \right).
\eeq
Four gauge conditions, $G_a = 0$, $a=1,...4$, are required---one for each
generator of \gwk.

\topic{Equivalence}
The demonstration of the equivalence between eqs.~\lowbrowint\ and
\holygrailint\ proceeds along lines that are similar to those used in the
abelian toy example presented previously. As was the case in this earlier
example, the equivalence works term-by-term in the lagrangian. The
correspondence between the field variables is
\eq
\label\correspond
\Sca_\mu \leftrightarrow A_\mu, \quad
\Scz_\mu \leftrightarrow  Z_\mu, \quad
\Scw_\mu^\pm \leftrightarrow W^\pm_\mu, \quad
\psi' \leftrightarrow \psi.
\eeq

The equivalence is explicit in unitary gauge, which is defined in this case
by the condition $\varphi^a(x) \equiv 0$, or equivalently $\xi(x) \equiv
1$, throughout spacetime. As is seen from the transformation rules of
eq.~\realization\ this condition does not completely fix the gauge. It is
preserved by the unbroken electromagnetic transformations which satisfy
$G = H = e^{i\omega_{\rm em}}$. In this gauge the relations for $\Scz_\mu$,
$\Scw_\mu$ and $\psi$ indicated in eqs.~\correspond\ above simply become
equalities.

More formally, using the unitary-gauge condition to perform the functional
integral over $\xi$ in eq.~\holygrailint, gives the result
\eq
\label\holygrailtwo
Z_2 = \int [d\Bfw_\mu] \; [d\psi] \;  \exp\left[
i\int d^4x \; \Scl_2 \right] \; \delta\left[G_\em \right] \;
\Det\left(\frac{\delta G_\em}{\delta \omega_\em} \right)  \;
\left.\Det \left( \frac{\delta\varphi^a}{\delta\omega^b} \right)
\right|_{\varphi=0}.
\eeq
Since $\Scl_2(\xi =1) = \Scl_1$ this clearly agrees with eq.~\lowbrowint\
apart from the final Fadeev-Popov-DeWitt ghost determinant that is
associated with the choice of unitary gauge
\eq
\delta\varphi^a(x)/\delta \omega^b(x') \equiv {\Delta^a}_b(x) \;
\delta^4(x-x').
\eeq

The final point is that the identity $\Det \equiv \exp \Tr \Log$ may be
used to rewrite this determinant as the exponential of a local, Lorentz-
and $U_{\rm em}(1)$-invariant function. As such it may be considered as a
shift in the parameters appearing in the original lagrangian, $\Scl_2$.
Furthermore, since  its contribution to $\Scl_2$ is proportional to
$\delta^4(x=0)$ its coefficients  are ultraviolet divergent and so their
contribution may be absorbed into the renormalizations that are anyhow
required in defining the functional integral of eq.~\holygrailtwo. In
practice the Fadeev-Popov determinant does not in any case arise until at
least two-loop order.

The practical benefit of this equivalence is that it allows the use of the
most convenient gauge for any particular application. Covariant gauges,
such as Feynman gauge, are particularly useful for making power-counting
arguments, since all propagators explicitly vary like $1/p^2$ for large
four-momenta, and the pathologies of the unitary-gauge propagator are put
into derivative couplings. For instance, this is the simplest way to
understand why QED remains renormalizable once a photon mass term is added,
while the same is not true for a nonabelian gauge theory. This distinction
is most easily seen from the form of the Nambu-Goldstone boson couplings.
While an invariant renormalizable lagrangian exists for a $U(1)$
Nambu-Goldstone boson --- \ie\ it is simply its kinetic term $-\hf \, D_\mu
\varphi D^\mu \varphi$ --- the same is {\it not} true for a nonabelian
symmetry group. This is because the kinetic terms are in this case not by
themselves invariant with respect to the nonlinearly-realized symmetries.
Conversely, unitary gauge has the simplicity of just involving physical
particles, allowing a direct identification of the physical significance of
the effective interactions.

\subsection{Derivative \vs\ Yukawa Couplings}

In this section we wish to make the previous arguments concrete by
considering an explicit one-loop example. Besides having applications later
in the paper, the example also serves to bring out three general, but
not-so-widely appreciated, features of the equivalence we have described.
These general points are listed at the end of the section.

It is a basic feature of the chiral lagrangian described above that all of
the would-be Nambu-Goldstone bosons (WBGB's) couple derivatively to all
other fields (and to themselves). This expresses a completely generic
feature of any Nambu-Goldstone-boson interaction, and is easily seen from
the expansion of \eg\ $\Scw_\mu(\xi)$ (\cf\ eqs. \auxiliary\ and
\connection) in powers of fields:
\eq
\label\wpowers
\Scw^\pm_\mu = W^\pm_\mu  - \frac{1}{\mw} \partial_\mu \varphi^\pm +
\cdots .
\eeq

The second term in this expansion gives a very simple Feynman rule for WBGB
couplings: simply contract the result for the corresponding $W^\pm$
coupling by $ik^\mu/\mw$, where $k^\mu$ is the WBGB four-momentum. This is
equally true regardless of whether the particle to which the $W^\pm$ or
$\varphi^\pm$ couples is a scalar, fermion or a gauge boson.

Notice that this type of coupling is {\it not} the same as what is obtained
for the WBGB's in a covariant gauge in the standard model. In the Standard
Model, for example, the WBGB--fermion interactions do not involve any
derivatives at all, since they come from the Yukawa couplings to the Higgs
multiplet. In the standard model these two formulations are physically
equivalent, since it is possible to pass from one to the other by
performing an appropriate field redefinition. As we shall now see, however,
they can and do give rise to different types of divergences in off-shell
quantities like the effective lagrangian. This point will become important
once we begin trying to track the cutoff dependence of loops in later
sections.

Consider then, the following effective interaction:\foot\gfourzft{This
interaction happens to violate CP and corresponds on shell to the
interaction denoted $g_4^Z$ in Ref.~\wwv.}
\eq
\label\anapole
\Scl_a = - a \; Z_\mu \left( W_\nu^+ W^{-\mu\nu} + W_\nu^- W^{+\mu\nu}
\right).
\eeq
The couplings that this interaction induces for the WBGB's are found by
substituting $W \to \Scw$ and $Z \to \Scz$ from eqs.~\connection, and
expanding the result in powers of fields. We choose to compute the
following CP-violating $Z$--$\tau$--$\tau$ vertex (or \zdm):
\eq
\label\zdmdef
\Scl_{\rm zdm} = - {iz \over 2}\; \ol{\tau} \; \gamma_5 \sigma^{\mu\nu} \tau \;
Z_{\mu\nu}.
\eeq
that this coupling induces at one loop.

In the unitary-gauge formulation we must evaluate the graph of
\fig\zdmloopfig\ Fig.~\zdmloopfig. The $W$ propagator that appears in each
of the two internal boson lines is:
\eq
\label\mvprop
G^{\mu\nu}_U(k) = \frac{-i}{k^2 - \mw^2}
\left[g^{\mu\nu} - \frac{k^\mu k^\nu}{\mw^2}\right].
\eeq

On the other hand, working with the chiral lagrangian in a general
covariant gauge leads not only to to Fig.~\zdmloopfig, but also to the
three other graphs that are obtained from this one by replacing each $W$
line by the corresponding WBGB propagator. In the standard
one-parameter-family of covariant gauges, the two types of boson
propagators that appear are:
\eqa
G_{(\eta)}(k) &= \frac{i}{k^2 - \eta \mw^2} \eolnn
\label\proprelation
{\rm and}\quad G^{\mu\nu}_{(\eta)}(k) &= \frac{-i}{ k^2 - \mw^2} \;
\left[ g^{\mu\nu} + (\eta -1) {k^\mu k^\nu\over k^2 - \eta \mw^2 }
\right]. \eeol
\eeq

The equivalence theorem of the previous sections argues that the
unitary-gauge result equals the sum of the four covariant-gauge graphs.
This is easy to see by using the following identity in the unitary-gauge
result:
\eq
\label\utoxiidentity
G^{\mu\nu}_U(k) =  G^{\mu\nu}_{(\eta)}(k)  + \frac{k^\mu k^\nu}{\mw^2} \;
G_{(\eta)}(k).
\eeq
The two factors $(ik^\mu/\mw) (-ik^\nu/\mw)$ are just what is required to
reproduce the Feynman rules for the WBGB couplings as given in
eq.~\wpowers. (The relative sign arises because momentum in at one end of
the boson line corresponds to momentum out at the other.) Thus, the
diagrams with WBGB's simply cancel the $\eta$-dependence of the $W$
propagators in Fig.~\zdmloopfig. Notice that it is crucial for this result
to use the derivative WBGB couplings for both the $WWZ$ vertex, {\it and}
the $W$-fermion vertices.

Since the integrands in the two formulations are equal, they give the same
result for the $\tau$ weak dipole moment, regardless of how the graphs are
regularized. We choose here to regulate the graph by inserting the form
factor, $F(p,\Lambda) = -\Lambda^2/(p^2 - \Lambda^2)$, into each internal
line.\foot\eachlineft{The momentum flowing through each line must be
regulated separately, or else the result will depend on how momentum is
routed through the graph.} (This may be viewed as a higher-derivative
regularization, for which a higher derivative kinetic term has been added
to the unperturbed lagrangian for each field.) The result for the
most-divergent part becomes:
\label\answercutoff
\eq
z_{\eightrm most-div} = - {a g^2\over 2304 \pi^2 }
\;\frac{\Lambda^2}{\mw^4} m_\tau,
\eeq
where $g$ is the $SU_{\sss L}(2)$ coupling constant.

It is instructive to compare this result with what would have been obtained
if we had used the chiral lagrangian only for the $WWZ$ vertex in
Fig.~\zdmloopfig, and had simply used Standard-Model Yukawa-type Feynman
rules for the WBGB-fermion vertices. In this case the WBGB graphs are less
divergent, since there are fewer powers of momentum associated with each
vertex. The most divergent part of the result becomes
\label\answerfeynman
\eq
z_{\eightrm most-div} = - {a g^2 \over 384 \pi^2} \;
{m_\tau \left(m_\tau^2 - m^2_{\nu_\tau}\right) \over \mw^4}
\; \ln \left({\Lambda^2 \over \mw^2 }\right).
\eeq

There are three lessons to be learned from this section and from this example:
\topic{1}
First, we explicitly verify the equivalence between the chiral lagrangian
and the lagrangian which ignores all but \gem. This equivalence relies
crucially on the derivative couplings of the WBGB's in chiral perturbation
theory. Criticizing the apparent non-gauge invariance of the TGV
lagrangians that are used in loop calculations (or, equivalently, of
unitary gauge) in favour of chiral perturbation theory clearly misses the
point. If there are problems with the large loop estimates that have been
obtained, then the reason must be found elsewhere. We point this reason out
in the following section.
\topic{2}
Next, we see explicitly that even the dominant cutoff dependence of
off-shell quantities, such as couplings in the effective lagrangian, depend
strongly on the choice of field variables used. In particular, the two
kinds of Feynman rules for the fermion--WBGB vertex may be obtained from
one another by performing a WBGB-dependent nonlinear field redefinition on
the fermion fields of the form $\psi \to f(\varphi) \psi$. (In fact, the
answer would have remained unchanged if the higher-derivative terms which
implement the cutoff were also transformed, since this transformation
introduces new cutoff-dependent fermion--WBGB interactions.) This is part
of the occupational hazard of trading in off-shell divergences: they depend
in detail on which field variables are regulated.
\topic{3}
But the last, and most important, point is this: without knowing the
underlying physics, which of these two answers is correct? If one
interprets $\Lambda$ to agree, in order of magnitude, with the new physics
scale, they have very different physical implications. The difference
between them could well be the difference between detecting $z$ at LEP or
not. We shall argue in the following sections that in this case it is
eq.~\answerfeynman\ which is correct. It is clearly important to be able to
decide which is right in advance!
\endtopic

\section{Cutoffs -- General Arguments}

The main point of this paper is to critically reassess the common habit of
inferring heavy-mass dependence from the cutoff dependence obtained purely
within low-energy loops. In this section we make our main points. In order
to do so, we start by presenting the arguments in favour of using cutoffs
in this way, followed by our criticisms of these arguments. We then provide
a few explicit examples to illustrate the relevant points.

\subsection{Why Might One Think That Cutoffs Track New Physics?}

Associating a physical interpretation with the cutoff is an almost
irresistable impulse when dealing with effective lagrangians. After all,
the effective theory is from the start only meant to describe physics below
some scale, $\Lambda$, above which we cannot probe. Since effective
theories are not renormalizable in the traditional sense, the insertion of
effective vertices into loop graphs can produce very divergent results. It
is natural to suppose that these divergences indicate that the amplitude in
question gets its most important contributions from the highest
frequencies: those just below the cutoff. Presumably this strong
sensitivity is removed once all of the heavy degrees of freedom of mass $M
> \Lambda$ are included, such as would happen if this underlying theory
were renormalizable. As a result, so the argument goes, the strong
short-distance contributions should saturate at $M$, leaving a result whose
size is set by replacing $\Lambda$ with $M$, at least up to order of
magnitude.

This reasoning can be made considerably more precise by rephrasing it as the
following principle
\ref\polchinski{J. Polchinski, private communication.}
\polchinski:
\smallskip{\narrower
If there is a divergent graph in the low energy theory, cutting it off at
the scale where the theory breaks down due to new physics gives a {\it
lower bound} to the actual value of the graph in the full theory (in the
absence of fine tuning).}
\smallskip

What could possibly be wrong with such a physically appealing argument? The
answer is that, in certain circumstances, nothing goes wrong with it.
Unfortunately, it can sometimes also happen that it is completely false,
and it fails because it does not take into account cancellations that are
automatically built into any effective theory. We describe here what these
cancellations are, and return in following sections to the question of how
to tell when the above reasoning will fail.

\subsection{The Curse of Cancellations}

Consider, then, a theory which involves two very different mass scales $M
\gg m$. (We have in mind that $m$ represents the weak scale --- say $m \sim
M_{\sss W}$ --- while $M$ represents the scale of unknown new physics.)
Suppose that within this theory we wish to compute a physical low-energy
observable, such as a calculable low-energy mass shift, $\delta \mu^2$, as
a function of these two mass scales. An example of this type of observable
in the electroweak interactions would be the deviation from unity of the
$\rho$-parameter, which is related to the comparative strength of the
low-energy charged- and neutral-current weak interactions.

We are interested in the form taken by $\delta \mu^2$ in the limit where
$m/M$ is taken to be asymptotically small, with all dimensionless
couplings held fixed. It is possible to make fairly general statements as
to the result in this limit (in four spacetime dimensions) if the
renormalizable part of the low-energy theory is perturbative, so that all
fields scale approximately as the noninteracting lagrangian would indicate.
Typically the answer in this case takes the following general form
\label\fullresult
\eq
\delta \mu^2(m,M) = c_0 M^2 + c_1 m^2 + c_2 \frac{m^4}{M^2} + \cdots
\eeq
in which the dots represent terms that are suppressed by more than two
powers of $m/M$. The dimensionless coefficients are functions of the other
(renormalized) dimensionless parameters of the theory, and they may also
depend at most logarithmically on the large mass ratio $M/m$. Notice that
the largest power of $M$ here is just set by dimensional analysis.

For applications to the electroweak interactions $m \sim \mw$, it is
important to be aware that the above form strictly applies only
asymptotically for $\mw/M \to 0$. It may therefore be expected to hold when
the new physics can be at very high scales compared to the weak scale, such
as if the underlying physics were a Grand Unified Theory of some kind. Its
application is less straightforward when the new physics is associated with
electroweak symmetry breaking, since in this case $M$ cannot be larger than
of order $4 \pi v$, and so $M/\mw \lsim 8\pi/g$. In this case eq.~\fullresult\
must be interpreted as applying to the $g \to 0$ limit rather than for $\mw/M
\to 0$ with $g$ fixed.

Imagine now performing the same calculation, but this time dividing the
contributions into a `low-energy' part and a `high-energy' part. To this
end choose a cutoff, $\Lambda$, which satisfies $m \ll \Lambda \ll M$.
First integrating out the high energy part of the spectrum produces a
low-energy effective lagrangian that is applicable at scales below
$\Lambda$. Next compute the physical mass shift in this low-energy
effective theory. Since this simply corresponds to a particular way of
organizing the calculation in the full theory it must produce the correct
answer of eq.~\fullresult\ above. The full expression may therefore be
broken up as follows
\eq
\delta \mu^2(m,M) = \delta \mu^2_{\rm le}(m,\Lambda,M)
+ \delta \mu^2_{\rm he}(m,\Lambda,M)
\eeq
in which the first (second) term here respectively contains only the
low-energy (high-energy) contributions. (This split between low and high
frequencies may be conveniently formulated in euclidean signature according
to whether the four-momentum, $p$, for a particle of mass $m_i$ in each
internal line of a Feynman graph satisfies the condition $p^2 + m_i^2 >
\Lambda^2$).

The low-energy and high-energy contributions to $\delta \mu^2$ in general
take the following form:
\eqa
\label\heresult
\delta \mu^2_{\rm he} &= c_0 M^2 + b_1 \Lambda^2 + \cdots \eolnn
\delta \mu^2_{\rm le} &= b'_1 \Lambda^2 + \cdots \eeol
\eeq
In both of these equations the ellipses represent terms that depend
differently on the small mass ratios $m/\Lambda$, $\Lambda/M$ or $m/M$ than
the terms that are explicitly written. Examples in later sections include,
for example, such quartically-divergent terms as $\Lambda^4/m^2$. Clearly
the condition that the two contributions sum up to the full result of
eq.~\fullresult, whose $\Lambda$-independence is manifest, requires that
the coefficients satisfy $b_1 + b_1' = 0 $, \etc

Now comes the main point. In order to calculate the scale of new physics
that may be probed by a detailed measurement of a quantity like $\delta
\mu^2$ we require the accurate knowledge of the coefficient $c_0$ in
eq.~\fullresult. If we only have access to the low-energy effective
lagrangian below scale $\Lambda$ then it is impossible to precisely compute
$c_0$. In particular, knowledge of the coefficient, $b'_1$, of the
low-energy quadratic divergence gives no \apriori\ information regarding
$c_0$, since it is completely cancelled by the high-energy contribution (or
counterterm) $b_1$. There is nothing miraculous about this cancellation; it
simply reflects how physics cannot depend on the intermediate steps in a
calculation.

There are occasions, however, when knowledge of the coefficient of a
particular divergence in the low-energy theory can be parlayed into
reliable information about the heavy-mass dependence of the full result
\ref\logtrick{S. Weinberg, in the proceedings of the Int. School of Subnuclear
Physics, Ettore Majorana, Erice, Sicily, (1976);
C.P. Burgess and A. Marini, \prd{45}{92}{R17}.}
\logtrick. A logarithmic divergence furnishes perhaps the simplest example.
Here the full and partial results for a dimensionless observable, call it
$A$, can take the following form
\eqa
\label\logcutoff
A & = A_{\rm le} + A_{\rm he} = a_0 \; \log\left( \frac{M^2}{m^2} \right) +
\cdots \eolnn
{\rm while} \quad A_{\rm he} &= a_0' \; \log\left( \frac{M^2}{\Lambda^2}
\right) + \cdots \eol
{\rm and} \quad A_{\rm le} &= a_0'' \; \log\left( \frac{\Lambda^2}{m^2}
\right)  + \cdots \eeolnn
\eeq
In this case the condition that the cutoff dependence cancel requires that
$a_0 = a_0' = a_0''$ and so the coefficient of the large logarithm within
the full theory may be determined simply by identifying the coefficient of
the logarithmic divergence within the low-energy theory. It is important to
realize that this property is {\it not} generically shared by other types
of divergences.

\subsection{``Good'' \vs\ ``Bad'' Variables}

With the general concepts regarding cutoffs now firmly in hand, we can now
demonstrate the flaw in the principle enunciated earlier (Section 3.1),
which states that cutoffs furnish lower bounds for the contributions of new
physics. A brief example here is instructive.

Consider the case where the standard model itself --- Higgs and all --- is
the low-energy theory, as might be appropriate to a Grand Unified Theory.
In this case  the bounds of eq.~\fullresult\ should apply since the
low-energy theory is perturbative in the regime of interest, and we may
take $\mw/M$ to be extremely small. Suppose we choose to compute the
cutoff dependence in this theory of the coefficient of the effective
operator $F_{\mu\nu} \sq F^{\mu\nu}$, which contributes to the vacuum
polarization of the photon. Since the standard model is renormalizable,
this result in a manifestly renormalizable gauge is finite --- varying like
$1/\Lambda^2$ for $\Lambda \gg \mw$. If the same result is computed in
unitary gauge, however, (or, equivalently, in any gauge using the
derivative WBGB couplings of chiral perturbation theory) then it diverges
quadratically: $\sim \Lambda^2 /\mw^4$. If taken seriously, this example
would drastically overestimate the heavy-mass dependence of the underlying
theory, which cannot be larger than $O(1/M^2)$.

Once again, just as in our earlier example involving the weak dipole moment
of the $\tau$, a change of variables has dramatically altered the
cutoff dependence of the effective lagrangian. These two examples
illustrate the difference between what might be called ``good'' and ``bad''
variables. To see the distinction between these variables, notice that for
both examples the divergences of the $S$-matrix are the same in {\it both}
sets of variables, since the $S$-matrix is unchanged by field
redefinitions. ``Bad'' variables are therefore characterized by large
cancellations in physical quantities, such as the $S$-matrix, between
enormous terms in the effective lagrangian. As a result, these are
variables for which the couplings in the lagrangian do not follow the
couplings that would be defined in terms of scattering amplitudes.

With this in mind, one can propose a modification of the above principle
\polchinski:
\smallskip{\narrower
If there is a divergent graph in the low energy theory, cutting it off at
the scale where the theory breaks down due to new physics gives a {\it
lower bound} to the actual value of the graph in the full theory (in the
absence of fine tuning), so long as ``good'' variables are used in the
calculation.}
\smallskip
It appears that this principle holds in all known examples. However, its
utility relies on the existence of an practical algorithm for determining
in advance whether the variables of interest are ``good'' or ``bad''.

This is our main criticism of the papers in Ref.~\errors. In using cutoffs
to regulate divergent loops involving anomalous TGV's, they obtain limits
on the coefficients of these operators which depend on their choice of
variables. Without knowing whether these variables are ``good'' or ``bad'',
one cannot ascertain if the bounds obtained are reasonable. A second
criticism of some of these papers, and indeed of some of those in
Ref.~\gbloopnoninv, is that the scale of new physics, $M$, is often allowed to
be greater than $4\pi v$, which is not permitted if the symmetry is realized
nonlinearly. This typically leads to overly stringent bounds on new operators.

\section{Banishing Cutoffs}

Rather than searching for a practical algorithm for ``good'' and ``bad''
variables, we prefer to recast the above principle in a way which does not
refer to cutoffs at all. It amounts, in essence, to the judicious use of
dimensional analysis, together with any other information that may also be
available purely within the low-energy theory. This information is all that
is really required of any analysis of low-energy graphs, and in
applications where cutoff dependence happens to track the underlying
masses, produces identical answers. It has the conceptual advantage,
however, of being insensitive to field redefinitions, and so of never
leading one badly astray through the mistaken use of ``bad'' variables. In
the remainder of this section, we describe this procedure, followed
immediately by a detailed calculation using a known model of underlying
physics with which both cutoff and our results can be compared.

Suppose, then, that some physics that is associated with a heavy mass
scale, $M$, (which might, for example, denote the mass of the lightest
unkown particle) is integrated out to produce a low-energy effective
lagrangian, $\leff$:
\label\operatorsum
\eq \delta \leff = c_n \Sco_n. \eeq
We are interested in the $M$--dependence of the coupling for an effective
operator of scaling dimension (mass)${}^{d_n}$ which appears in this
effective lagrangian. In general this is an ill-defined question, since the
dependence of $c_n$ on heavy physics requires a proper definition of the
composite operator it multiplies, $\Sco_n$. We therefore pause here to make
a brief aside concerning a particularly convenient formalism for these
purposes.

\subsection{A Regularizational Aside}

A particularly clean and convenient scheme with which to work in an
effective theory is dimensional regularization supplemented by the
`decoupling subtraction' renormalization scheme
\ref\decoupling{S. Weinberg, \plb{91}{80}{51};
B. Ovrut and H.J. Schnitzer, \prd{24}{81}{1695}; \npb{184}{81}{109};
L. Hall, in {\it TASI Lectures in Elementary Particle Physics}, edited by
D. Williams, (Ann Arbor, 1984); \npb{178}{81}{75};
F. Gilman and M. Wise, \prd{27}{83}{1128};
R. Miller and B. McKellar, \prc{106}{84}{170};
I. Hinchliffe, in {\it TASI Lectures in Elementary Particle Physics};
H. Georgi, {\it Weak Interactions and Modern Particle Theory}
(Benjamin/Cummings Menlo Park, 1984); \npb{363}{91}{301}.}
\decoupling. This scheme consists of minimal subtraction supplemented by
the explicit removal of heavy degrees of freedom as the renormalization
point is lowered below the corresponding mass thresholds. This `integrating
out' of the heavy particles is in practice implemented as a set of matching
conditions for the appropriate effective couplings at these threshholds.
The resulting couplings may then be used as initial conditions for the
renormalization group equations that define the scale-dependence of such
couplings in the theory below the threshhold. With this scheme a
logarithmic dependence on the masses of the problem (including $M$) is
introduced into the coefficients $c_n$ as the various effective operators
are evolved between particles threshholds.

The beauty of using dimensional regularization in this way is that no
confusion is possible between the cutoff and the heavy-physics scale, since
within this framework no cutoff, $\Lambda$, arises at all. As a result only
the physical masses ever arise in effective couplings. Furthermore, more
and more divergent graphs in the effective theory, which involves only
light particles, simply introduce higher and higher powers of the light
mass, $m$, rather than some higher scale such as $\Lambda$ or $M$. As a
result it becomes possible (and convenient) to include within the loops of
the low-energy theory {\it all} of the momenta of the light fields, right
up to infinity. This leads to a real distinction in the nature of the
matching between the underlying theory and the effective theory when using
cutoffs and dimensional regularization. When using a cutoff, all
frequencies above the scale $\Lambda$ are integrated out, including all of
the modes of the heavy particles as well as the high frequency components
of the light particles. In dimensional regularization, one instead
integrates out {\it only} the heavy-particle contributions; leaving all of
the momenta of the light fields in the low-energy theory. This allows the
matching between the effective and the underlying theories to be made at
the heavy mass threshold itself, and so the only mass which appears due to
this matching is typically this threshold mass, $M$.

There is another practical benefit in using dimensional regularization.
Dimensionally-regularized graphs are much less sensitive to the field
redefinitions that relate the ``good'' and ``bad'' variables of the earlier
examples. For instance, if dimensional regularization is used to regularize
the contribution of the $WWZ$ interaction to the \zdm, the divergent piece
is found to be
\eq
\label\answerdimreg
 z_{\eightrm pole} = {ag^2\over 384\pi^2}
\, {m_\tau \left(m_\tau^2 - m^2_{\nu_\tau}\right) \over M_W^4}
\,  \left({1\over \epsilon}\right).
\eeq
where $n = 4 - 2\epsilon$ is the dimension of spacetime. This result holds
using {\it either} Yukawa-type or derivative couplings for the WBGB's to
the fermions.

Within minimal subtraction, we find therefore that the renormalized
parameter $z$ mixes with the renormalized parameter $a$ in the following
way:
\eq
\label\running
z(\mu) = z(M) - \; {g^2\over 384\pi^2} \,
{m_\tau \left(m_\tau^2 - m^2_{\nu_\tau}\right) \over M_W^4} \; a(M) \;
\ln \left(  \frac{M^2}{\mu^2} \right).
\eeq
Notice the similarity between the logarithmic dependence here, and the
previous results of eq.~\answerfeynman.

Both terms in eq.~\running\ have a clear interpretation. The logarithmic
dependence corresponds to the explicit operator mixing that can be
unambiguously computed purely within the low-energy effective theory. The
initial conditions, $z(M)$ and $a(M)$, however are determined by matching
to the underlying theory and so cannot be known until this theory is
specified. At best we can only try to estimate the size of these initial
conditions, and this is the goal of the remainder of this section.

\subsection{The Generic Estimate}

With this definition in mind, we wish now to estimate how the couplings
$c_n$ of eq.~\operatorsum\ depend on the new-physics scale, $M$. We are
specifically interested here in the powers of $M$ that arise at the
threshhold, $M$, rather than any logarithmic dependence. Simple dimensional
analysis would indicate:
\label\dimanal
\eq  c_n = \hat{c}_n \; M^{4-d_n}. \eeq
Without any additional information about the nature of the new physics that
is responsible for this effective lagrangian, all that can be said about
the dimensionless coupling, $\hat{c}_n$, is that it is $O(1)$ {\it or
smaller}.

With more assumptions concerning the physics at $M$, more information can
be extracted about the $c_n$. We next illustrate how different kinds of
physics can differ in their implications for $c_n$ by contrasting two
plausible alternatives for electroweak symmetry-breaking physics at $M$.

\vfill\eject
\subsection{Strong Coupling: Naive Dimensional Analysis}

Suppose first that the symmetry-breaking sector is strongly coupled, with
only the WBGB's appearing at energies much less than $M$. In this case
chiral perturbation theory organizes their couplings according to the
numbers of derivatives which appear in the lagrangian. For applications to
energies that are much less than the electroweak scale, $v$, simple
dimensional analysis with $M \sim v$ properly describes the size of each
interaction.

Of more practical interest, however, is the application of this lagrangian to
electroweak energies, $E \simeq v \ll M$. In this case higher-derivative
interactions should be suppressed by powers of $M$ rather than $v$, and it
becomes important to keep track of the powers of $v/M$ which can appear in
the coefficients $c_n$. A set of self-consistent statements for the sizes
that can be expected for any given term in the chiral lagrangian is called
``Naive Dimensional Analysis'' (NDA)
\ref\nda{A. Manohar and H. Georgi, \npb{234}{84}{189}; H.Georgi and
L.Randall, \npb{276}{86}{241}; H. Georgi, \plb{298}{93}{187}.}
\nda. It states that a term having $b$ WBGB fields, $f$ weakly-interacting
fermions fields, $d$ derivatives and $w$ gauge fields has a coefficient
whose size is:
\label\ndaestimate
\eq  c_n(M) \sim v^2 M^2 \; \left( { 1\over v} \right)^b \; \left(
{ 1\over M^{3/2}} \right)^f \; \left( { 1\over M} \right)^d \;
\left( { g\over M} \right)^w,
\eeq
with $M \lsim 4 \pi v$. (If the fermions are strongly interacting, then the
appropriate factor is $1/v\sqrt{M}$ for each fermion.)

Some examples of this counting are instructive, particularly when these are
compared with the alternative estimates of the next section. For instance,
according to the above estimate, the mass terms for the $W$ and $Z$ bosons
are both of order $g^2 v^2$. This indicates that the small size of the
deviation from unity of the rho parameter cannot be understood in this
picture as being simply the result of a suppression by powers of $v/M$.
Additional approximate symmetries are required in order to explain the
small size of $\delta \rho$. Also, typical corrections to the charged- and
neutral-current interactions for fermions are here of the order of
$gv^2/M^2$. Finally, triple-gauge boson operators such as $\kappa W^*_\mu
W_\nu Z^{\mu\nu}$ and $\lambda W^{*\mu \nu} W_{\nu \lambda}
{Z^\lambda}_\mu$ are respectively of order $\kappa \sim g^3 v^2/M^2$ and
$\lambda \sim g^3 v^2 / M^4$. We next compare these estimates with the
implications of an alternative scenario.

\subsection{Weak Coupling: Linearly-Realized Lagrangian}

An alternative perspective arises if the low-energy theory fills out a
linear realization of the electroweak group. In this case $M$ need not be
small compared to $4\pi v$, and the WBGB's fall into some linear
representation of this group. Again operators can have coefficients that
are suppressed by powers of $v/M$ once the low-energy Higgs fields are
given their {\it v.e.v.}s, and so the power that arises depends on the
representation in which the symmetry-breaking order parameter transforms.
Much the most plausible choice for such a linearly-realized Higgs
representation is one or more doublets, with the standard hypercharge
assignment. In this case the dependence on $v/M$ of any non-Higgs
interactions may be found by taking $c_n = \hat{c}_n M^{4 - d_n}$, as
before, and then replacing any Higgs multiplets in the effective operator
by their {\it v.e.v.}s. In this case the linearly-realized gauge symmetry
enforces relations amongst the coefficients of operators of a given type,
depending on how these operators fall into linearly-realized multiplets.

This is best illustrated with a few examples. Consider the $W$ and $Z$
boson mass terms: $\Sco_{\sss W} = W^*_\mu W^\mu$ and $\Sco_{\sss Z} = \hf
\, Z_\mu Z^\mu$. The lowest-dimension operator which contains these terms
is simply the dimension-four Higgs kinetic term, $(D_\mu \phi)^\dagger
(D^\mu \phi)$. Just as for the standard model, replacement of $\phi$ by its
expectation value in this operator generates the particular combination
$\cos^2 \theta_w \, \Sco_{\sss W} +  \Sco_{\sss Z}$ with a coefficient that
is of order $g^2 v^2$. More general combinations arise at dimension six,
such as through the operator $(\phi^\dagger D_\mu \phi) \, (\phi^\dagger
D^\mu \phi) / M^2$. This and similar operators ruin the mass relation $\mw
= \mz \cos\theta_w$, by amounts that are of order $g^2 v^4/M^2$. In
contrast with the NDA estimate, $\delta \rho$ is automatically small if
$v^2/M^2 \ll 1$.

As we shall see in a later section, the smallness of the present estimate
in comparison with the NDA result has a simple explanation within the
context of an underlying multi-Higgs  model. In this case contributions
such as those to $\delta \rho$ typically arise at one loop and are
proportional to $g \lambda_{\sss H}^2 / 16 \pi^2$, where $\lambda_{\sss H}
\simeq g \mh/\mw$ is a Higgs self-coupling. If this self-coupling is weak,
then the suppression by $1/(4 \pi)^2$ corresponds to a factor of $v^2/M^2$.
Once $\lambda_{\sss H} $ is of order $4 \pi$, however, for which $\mh \sim
4 \pi v$, this suppression is lost and we obtain the NDA result.

It is not always true that NDA gives a larger estimate for effective
couplings than would a linearly-realized underlying theory, however. For
example, both predict deviations from the standard model charged- and
neutral-current couplings that are of order $gv^2/M^2$. Similarly, both
estimates for the coupling $\kappa W^*_\mu W_\nu Z^{\mu\nu}$ are of order
$g^3 v^2/M^2$. Furthermore, for the coupling $\lambda$ (which premultiplies
the interaction $W^{*\mu \nu} W_{\nu \lambda} {Z^\lambda }_\mu$), the NDA
estimate is actually {\it smaller} than that for a linearly realized model.
NDA would predict $\lambda \sim g^3 v^2 /M^4$ while the linearly-realized
estimate is $\lambda \sim g^3 / M^2$, since this interaction can be
embedded into the linearly-realized operator $\Tr[ W^{\mu \nu} W_{\nu
\lambda} {W^\lambda}_ \mu]$ without the necessity for Higgs doublets.

\subsection{Using Loops to Infer Further Information}

These estimates that are simply based on dimensional analysis can be
sharpened using additional assumptions. A dimensional estimate for $c_n(M)$
can be obtained by using loops in the low-energy theory to estimate factors
of dimensionless coupling constants and $1/(16 \pi^2)$ which arise from the
low-energy contribution to $c_n$ at lower scales. If these are assumed to
not cancel with the high-frequency contribution, then these factors may be
used to place a {\it lower bound} on $c_n$, just as in the principle that
was enunciated in the earlier section to describe the potential relevance
of cutoff dependence. The main difference in the present formulation is the
use of these loops purely to determine the dependence on dimensionless
combinations of couplings, with only dimensional analysis being used to fix
the dependence on $M$.

\ref\georgi{For a somewhat related discussion regarding the use of field
redefinitions to eliminate derivatives, see H. Georgi, \npb{361}{91}{339}.}
For loops which involve WBGB's there is one dimensionless coupling that is
of  particular interest. This is the dimensionless coupling which describes
the interactions between WBGB's and the other particles of the theory. It
is always possible to choose variables such that these are proportional to
the ratio of the particle's mass to $v$, rather than using the derivative
coupling of Section (2). For instance: $\lambda_{\varphi ff} \sim g m_f /
\mw$, and $\lambda_{\varphi ww} \sim g \mw^2/v$, \etc. Including these
couplings is important if the corresponding particle masses are large, in
that they can produce what appears to be a positive power of a heavy mass.
We illustrate this in more detail in the following section. Use of this
coupling strength amounts to using our freedom to use field redefinitions
to  remove as many derivatives as possible from WBGB couplings \georgi.
This is where the use of `good' variables enters our rules \polchinski.

This procedure is clearly operationally very similar to what is usually
done when using cutoffs to estimate effective interactions. In particular,
it reproduces the many successful estimates that are often argued from
using cutoffs. The main difference is that the power of $M$ that
contributes here is  explicitly argued purely on dimensional grounds,
thereby removing the uncertainty that is associated with the choice
of ``good'' and ``bad'' variables.

\section{Known New Physics: An Example}

We now wish to apply this reasoning to a model for which all of the
heavy-mass dependence is known and calculable. This permits a comparison of
the above arguments with the known correct dependence on $M$, as well as
with the cutoff dependence of the low-energy effective theory.

\subsection{An Explicit Calculation}

We consider a two-Higgs doublet model with soft CP-breaking terms in the
Higgs potential, and where we imagine that the physical Higgs particles all
have masses that are as large as is possible: $\mh \lsim 4 \pi v$. Larger
masses are not possible here without disbelieving the perturbative
analysis, since the Higgs masses can be made larger than $v$ only by
increasing their self couplings. In this model the anomalous $WWZ$ coupling
of eq.~\anapole\ arises at one loop, with a calculable coefficient. We may
therefore compute the contributions which this operator makes to the \zdm\
of section (2), as well as to the $\rho$-parameter, and contrast this with
an estimate of the corresponding higher-loop graphs that are obtained
within the underlying theory when the effective WWZ vertex is resolved.

\ref\he{X.G. He, J.P. Ma and B.H.J. McKellar, University of Melbourne
preprint UM-P-92/75.}
Following Ref.~\he, we consider a two-Higgs doublet model in which CP is
spontaneously broken. This occurs when there is a relative phase between
the vacuum expectation values (vevs) of the two Higgs doublets. In such a
scenario, tree-level flavour changing neutral currents (FCNC's) are usually
generated, but these can naturally be made small if CP violation is
generated via soft CP breaking terms in the Higgs potential. The two Higgs
doublets can then be written $\phi_i^{\sss T} = (\phi_i^+,\phi_i^0 + v_i
e^{i\theta_i}),~i=1,2$, in which $v_i e^{i\theta_i}$ are the vevs. For
calculational purposes, it is useful to change bases such that the WBGB
fields ($\varphi_{\sss Z}^0,\varphi_{\sss W}^+$) are decoupled from the
physical Higgs  fields ($H^+,H_{1,2}^0,I_2^0$). The new basis is
$\phi_1^{\prime{\sss T}} = (\varphi_{\sss W}^+,H_1^0 + i \varphi_{\sss Z}^0
+ \sqrt{v_1^2+v_2^2})$, $\phi_2^{\prime  {\sss T}} = (H^+,H_2^0 + i
I_2^0)$, in which only the vev of $\phi_1^{\prime}$ is nonzero. Although
$H^+$ is a mass eigenstate, the neutral states $H_{1,2}^0$ and $I_2^0$ are
not. They are related to the mass eigenstates by an orthogonal matrix
$d_{ij}$:
\eq
\pmatrix{H_1^0 \cr H_2^0 \cr I_2^0 \cr} =
\pmatrix{d_{11} & d_{12} & d_{13} \cr
         d_{21} & d_{22} & d_{23} \cr
         d_{31} & d_{32} & d_{33} \cr}
\pmatrix{\phi_{m1} \cr \phi_{m2} \cr \phi_{m3} \cr}.
\eeq
In the absence of CP violation, $d_{13}=d_{23}=d_{31}=d_{32}=0$.

\topic{The $WWZ$ Effective Operator}
There are two graphs which contribute at one loop to the CP violating $WWZ$
vertex in eq.~\anapole. These are shown in \fig\anapolefig\
Fig.~\anapolefig. In fact, since we are only interested in getting an idea
of the dependence on the Higgs' masses, we concentrate only on diagram (a)
in Fig.~\anapolefig.

If we had no knowledge of the underlying theory, we could estimate the
dependence of $a$, the coefficient of the $WWZ$ vertex, on the heavy mass
scale $M \sim \mh$ by using the dimensional analysis of the previous
section. In order to determine the suppression by $v$ that is appropriate,
we use the estimate of NDA, since this is appropriate to the case of a
strongly-coupled Higgs bosons that we are considering. Since $a$ is
dimensionless, we expect (after inserting a factor of $g$ for each vector
boson):
\label\expectwwz
\eq a_{\rm dim} \sim g^3 \left( {v^2 \over M^2} \right) \sim g
\left( {g \over 4 \pi} \right)^2 . \eeq
Direct calculation, on the other hand, gives
\eq
a_{\rm model} = \sum_{ij}\left\{i{g^3\over 32\pi^2\cos\theta_W}
\left( d_{3i}d_{2j} - d_{3j}d_{2i} \right)
\left( d_{2i}d_{2j} + d_{3i}d_{3j} \right) \left(m_i^2-m_j^2\right) I_1
\right\},
\eeq
where
\eq
I_1 = \int_0^1 dx_1 \int_0^{x_1} dx_2 x_2(x_1-x_2)
{1\over x_2(m_i^2-m_j^2) + x_1(m_j^2-m_c^2) +m_c^2 +\mw^2 x_1(x_1-1) }.
\eeq
In the above equations, the sum is over the physical neutral Higgs bosons
with $m_i$ and $m_c$ being the neutral and charged Higgs masses,
respectively. From the above expression, it is clear that
\label\orderofmag
\eq  a_{\rm model} \simeq  {g^3 \over 16\pi^2} \;
\ln \left( {\mh^2 \over \mw^2} \right),
\eeq
where $\mh$ is a generic Higgs mass. This agrees with the estimates from
dimensional analysis, for $\mh \sim M \lsim 4 \pi v$.

For our later purposes we wish to embed the anomalous $WWZ$ interaction
into loops in order to estimate their implications for other effective
interactions. Since the strongest dependence on heavy masses comes from the
longitudinal $W$ particles, $\varphi_{\sss W}$, in these loops, we pause
here to present an estimate for the size of the coefficient of the
anomalous $Z\varphi_{\sss W}\varphi_{\sss W}$ vertex. There is only one
difference from the previous case: the $\varphi_{\sss W}$ bosons couple
with a strength that is proportional to $gM/\mw$ rather than simply to $g$.
On dimensional grounds the largest contributions to the $Z \varphi_{\sss W}
\varphi_{\sss W}$ coupling should be proportional to:
\label\expectppz
\eq a^\varphi_{\rm dim} \sim g \left( {v^2 \over M^2} \right)\;
\left({gM \over \mw} \right)^2 \sim g .
\eeq

In this model, the lowest-dimension anomalous $Z \varphi_{\sss W}
\varphi_{\sss W}$ coupling arises at one loop from the graphs of
Fig.~\anapolefig. Keeping only terms linear in  $q$, the four-momentum of
the external $Z$, we find that the $Z\varphi_{\sss W} \varphi_{\sss W}$
vertex is
\eq
\sum_{ij}\left\{ -{g^3\over 4\cos\theta_W}
\left( d_{3i}d_{2j} - d_{3j}d_{2i} \right)
\left( d_{2i}d_{2j} + d_{3i}d_{3j} \right)
{\left(m_i^2-m_j^2\right) \left(m_i^2-m_c^2\right) \left(m_j^2-m_c^2\right)
\over \mw^2} I_2^\mu \right\},
\eeq
with
\eq
I_2^\mu = \int {d^4 l \over (2\pi)^4} {(2 l)^\mu (2q\cdot l) \over
\left[ (l+K/2)^2 - m_c^2 \right] (l^2-m_i^2)^2 (l^2-m_j^2)^2}~,
\eeq
where $K=k-k'$. It is not necessary to solve this integral exactly -- what
is important is that for the external momenta of interest (\ie\ those that
are $\lsim \mh$) it is dominated by momenta of order $\mh$, giving an
integral that is of order $\mh^{-4}$. This gives the result
\label\modelppw
\eq  a^\varphi_{\rm model} \simeq {g^3 \over 16\pi^2} \;
\left( {\mh \over \mw}\right)^2 \; \ln \left( {\mh^2 \over \mw^2} \right),
\eeq
which is larger by an additional factor of $\mh^2/\mw^2$ in comparison with
the result for transverse $W$'s: eq.~\orderofmag. This enhancement
corresponds, in the underlying theory, to the replacement of two gauge
couplings, $g$, with  two WBGB-Higgs couplings, $\lambda_{{\sss HH}\varphi}
\sim g\mh/\mw$. It agrees with estimate \expectppz\ when $ \mh \sim M \sim
4 \pi v$.

\topic{The Weak Dipole Moment}
Next consider the \zdm\ of eq.~\zdmdef\ in this effective theory. Using
only naive dimensional analysis, we can therefore only conclude:
\eq z_{\rm dim} \sim  {g v^2\over M^3}. \eeq
Any further information is more model specific.

In order to sharpen our estimate we next consider the size of the \zdm\
that is induced in the low-energy theory from the effective $WWZ$ operator
considered previously, \via\ the loop of Fig.~\zdmloopfig. The dominant
short-distance behaviour comes from the contributions of longitudinal $W$'s
to this graph. In this case there is now  an additional factor of $m_\tau$
from the required helicity flip, as well as two factors of the longitudinal
$W$ couplings to the fermion line, $\lambda_{\varphi \tau\tau} \sim g \,
m_\tau/\mw$. Taking our estimate for this graph as a lower bound, we
therefore expect:
\label\zdmestimate
\eqa  z_{\rm dim} &\gsim   {a^\varphi_{\rm dim} \over 16 \pi^2}  \;
\left( { g \, m_\tau\over \mw}  \right)^2  \; { m_\tau \over M^2} \eolnn
&\sim  {g^5\over (4 \pi)^2}  \; \left( {v^2\over \mw^2} \right) \; \left(
{ m_\tau^2 \over \mw^2 }\right) \; \left( {m_\tau \over M^2} \right), \eolnn
& \gsim { g^5 \over (4 \pi)^4} \; \left( { m_\tau^3 \over \mw^4} \right) .
\eeol
\eeq
We have used $v^2/M^2 \gsim 1/(4 \pi)^2$ in this last equation.

In the underlying theory, the \zdm\ appears at two loops \fig\twoloopzdm\
as in Fig.~\twoloopzdm. The strongest dependence on $\mh$ again comes when
both  $W$'s in the loop are longitudinal --- in a covariant gauge they are
WBGB's. Although the full 2-loop diagram is difficult to solve completely,
it is sufficient for our purposes to estimate the integrals using
dimensional analysis. Again, the important region in the loop integration
comes from momenta $\sim \mh$, since $\mh$ is the largest scale in the
problem. Including the factor of $m_\tau$ due to the required helicity-flip
on the fermion line, and two factors of the WBGB-$\tau$ Yukawa coupling:
$\lambda_\tau \simeq g m_\tau/\mw$, we arrive at the following estimate for
$z$:
\label\zdmmodelcalc
\eq z_{\rm model} \sim g^5 \left( {1\over 16\pi^2} \right)^2 \;
{m_\tau^3\over \mw^4} \; \ln^2\left({\mh^2\over \mw^2}\right). \eeq
This result agrees both with the our current estimate of eq.~\zdmestimate,
as well as with the earlier cutoff-based estimate of eq.~\answerfeynman,
but {\it not} with the `bad-variable' result of eq.~\answercutoff.

\topic{The Vacuum Polarization}
It is instructive to also consider the contributions towards the $Z$ vacuum
polarization that are induced by the $WWZ$ operator in this model. Besides
providing another comparison with the estimates, it furnishes an example
for which there is (superficially) an {\it enhancement} by powers of
$M/\mw$, and for which a simple cutoff analysis in unitary gauge proves to
be correct.

The required contribution to the $Z$ vacuum polarization comes from the
three-loop graph of \fig\vacpolfig\ Fig.~\vacpolfig. Again, in the
underlying model the largest contribution comes when both $W$'s in the
inside loop are longitudinal.  Then each $Z\varphi_{\sss W}\varphi_{\sss
W}$ vertex contributes a factor of order
\eq
a^\varphi_{\rm model} \; q, \eeq
where $a^\varphi_{\rm model}$ is given in eq.~\modelppw, and $q$ is the
four-momentum  which flows through the external $Z$ line. The middle loop
gives just the loop  factor $1/16\pi^2$ times a logarithm. Therefore we
find that the contribution to  the $Z$ vacuum polarization in this model
has the form
\label\zdminmodel
\eq
\left[ \delta\Pi_{\sss ZZ} \right]_{\rm model} \sim
\left({g^2 \over 16\pi^2}\right)^3 \; {\mh^4\over \mw^4} \; q^2 \;
\ln^3\left({\mh^2\over \mw^2}\right).
\eeq
Notice the large power of $\mh/\mw$. This result {\it agrees} with the
most-divergent  part of the unitary gauge cutoff dependence that is
obtained by inserting two effective $WWZ$ interactions into a one-loop
vacuum polarization diagram:
\eq
\label\zcutoff
\left[ \delta\Pi_{\sss ZZ} \right]_{\eightrm most-div} \left(q^2\right) =
- { a^2 \over 576 \pi^2 } \, { \Lambda^4 \over M_W^4 } \, q^2 . \eeq
if we take our earlier estimate for the $WWZ$ interaction: $a_{\rm dim}
\sim g^3  v^2/\mh^2 \gsim g^3/(4\pi)^2$.

Our dimensional estimate for this quantity, on the other hand, is
\label\vacpolest
\eqa \left[ \delta\Pi_{\sss ZZ} \right]_{\rm dim} &\sim {1 \over (4 \pi)^2} \;
(a^\varphi_{\rm dim})^2 \eolnn
& \sim {g^6 \over (4 \pi)^2} \; \left( {v^4 \over M^4} \right) \;
\left( {M^4 \over \mw^4} \right) , \eeol
\eeq
which also agrees with the result of the underlying model once we use
$v/M \gsim 1/4\pi$.

Notice that, keeping in mind $g \mh \lsim 8 \pi \mw$, what appears to be an
enhancement of four powers of $\mh/\mw$ in eq.~\zdminmodel\ is really more
than compensated for by the suppression by six powers of $g/4 \pi$, as it
must be in order for the result to be sensible. Thus, it is misleading in
this case to  use the corresponding enhancement in eq.~\zcutoff\ without
also including the accompanying suppression that is implicit in the
coefficient $a$.

\section{Conclusions}

Effective lagrangians are the natural way to parametrize the effects of the
new {\nobreak physics} which must lie beyond the standard model. The next
generation of experiments will have the ability to probe a number of these
new effective operators. Quite naturally, then, one wants to have an idea
of how big these new effects might be.

Much work has gone into constraining the new operators, particularly those
corresponding to trilinear gauge boson vertices, through their loop
contributions to quantities which are measured at lower energies. We have
argued here that these estimates \errors\ are typically misleading, and
often give bounds which are overly stringent.

Other authors \criticisms\ have made the same criticisms. However, they
trace the cause of the problem to the apparent non-gauge invariance of the
operators that are widely used in the literature. We argue instead that in
this instance gauge invariance is a complete red herring and is not the
source of the problem.

If one does not wish to explicitly include a Higgs scalar in the low-energy
theory, there are two principal candidates for such an effective lagrangian
-- one which requires only $U_\em(1)$ gauge invariance, but not
$SU_L(2)\times U_Y(1)$ gauge invariance, and one which imposes the full
$SU_L(2)\times U_Y(1)$ gauge invariance, nonlinearly realized. We have
demonstrated the equivalence of these two lagrangians.

The same arguments as are used here may be similarly used to prove this
equivalence for more general symmetry-breaking patterns $G \to H$. This
shows that any effective theory containing light spin-one particles
automatically has a (spontaneously broken) gauge invariance. Alternatively,
one can say that at low energies there is little to choose between a
spontaneously-broken gauge invariance and no gauge invariance at all.

The real source of the problems is the widespread use of cutoffs to
regulate divergent graphs in the low-energy effective lagrangian. Both the
effective lagrangian and its divergences, being off-shell quantities, are
not invariant under field redefinitions. As a consequence, the result of a
loop calculation will generically depend on the choice of variables, if
cutoffs are used to regulate the divergences.

It is in principle possible to use ``good'' variables in such loop
calculations, in which case the cutoff behaviour of the final answer
accurately reflects the true dependence of the operator on the heavy mass
scale $M$. However, it is equally possible to choose ``bad'' variables,
characterized by  cancellations in the S-matrix between large terms in the
effective lagrangian, in which case the cutoff does {\it not} properly
track the dependence on $M$. If ``bad'' variables are used, the bounds on
effective operators inferred from such calculations are typically much too
strong, and completely unreliable. In the absence of an algorithm to
distinguish ``good'' and ``bad'' variables, the constraints obtained from
such cutoff-regulated calculations are ambiguous at best.

A separate mistake that has also been made when bounding effective
interactions has been to take the scale of new physics $M$ to be 10 TeV, or
higher \gbloopnoninv, \errors, even when the effective theory does not {\it
linearly} realize the electroweak gauge group. In this case the effective
lagrangian is simply being applied beyond its domain of applicability,
since perturbative unitarity typically fails for such models when $M \gsim
4 \pi v$.

If one wants to estimate the size of the new operators, we advocate
dispensing with cutoffs completely. A simpler method is to just use
simple dimensional arguments, supplemented by any additional information
concerning dependence on coupling-constants and $(4 \pi)$'s that can be
gleaned by inspecting underlying or low-energy graphs. These rules coincide
in practice with currently-used lore when this lore is sufficiently well
spelled out. It has the conceptual advantage of not relying on the
cutoff dependence of low-energy diagrams.

One quantity which {\it is} accurately calculable within the low-energy
effective lagrangian (as opposed to being an order-of-magnitude estimate)
is the mixing among operators as the effective lagrangian is evolved down
from the heavy mass scale $M$ to low energies. This mixing, which is always
logarithmic, is most easily computed using dimensional regularization,
along with  the decoupling-subtraction  renormalization scheme. Among the
beauties of dimensional regularization is that it is comparatively
insensitive to the choice of ``good'' or ``bad'' variables.

By using dimensional regularization to calculate the mixing of operators,
and dimensional analysis to estimate the size of the initial conditions,
\ie\ the effective operators at scale $M$, one sees that it is never
necessary to deal with cutoffs in a low-energy effective lagrangian.

\bigskip
\noindent
Note Added:

After this paper was released, we have become aware of
\ref\einhorn{C. Arzt, M.B. Einhorn and J. Wudka, preprint NSF-ITP-92-122I.}
Ref.~\einhorn, whose authors present a point of view more similar to our
own.

\bigskip
\centerline{\bf Acknowledgments}
\bigskip

We would like to thank Joe Polchinski for numerous enlightening discussions
on the subject of cutoffs and effective lagrangians. D.L. thanks F. del
Aguila for the hospitality of the University of Granada, where part of this
work was done. Many thanks also to Fawzi Boudjema, Steven Godfrey, Markus
Luty, Ivan Maksymyk, Yossi Nir, Santi Peris, Xerxes Tata and German
Valencia for helpful criticism. This research was partially funded by funds
from the N.S.E.R.C.\ of Canada and les Fonds F.C.A.R.\ du Qu\'ebec.

\vfill\eject
\centerline{\bf Figure Captions}
\bigskip

\topic{Figure \zdmloopfig}
The Feynman graph through which the anomalous gauge-boson vertex
contributes to fermion weak dipole moments.

\topic{Figure \anapolefig}
The Feynman graphs which generate the CP-violating anomalous gauge-boson
vertex in the two-Higgs model.

\topic{Figure \twoloopzdm}
A Feynman graph which generates the CP-violating $\tau$ \zdm\ in the
two-Higgs model.

\topic{Figure \vacpolfig}
The 3-loop contribution to the $Z$-boson vacuum polarization. The blobs
indicate the 1-loop anomalous gauge-boson vertices whose structure is shown
in Fig.~\anapolefig.

\listrefs

\bye